\documentclass[conference]{IEEEtran}

\usepackage{graphicx}
\usepackage{epsfig}
\usepackage{epstopdf}
\usepackage{amssymb}
\usepackage{amsmath}
\usepackage{amssymb}
\usepackage{float}
\usepackage{algorithm}
\usepackage{algpseudocode}

\makeatletter

\newcommand{\Rmnum}[1]{\expandafter\@slowromancap\romannumeral #1@}
\makeatother

\ifCLASSINFOpdf
\else
\fi
\begin{document}
%
\title{A Cooperative Scheduling Scheme of Local Cloud and Internet Cloud for Delay-Aware Mobile Cloud Computing}

\author{
\IEEEauthorblockN{Tianchu Zhao, Sheng Zhou, Xueying Guo, Yun Zhao, Zhisheng Niu}
\IEEEauthorblockA{
Tsinghua National Laboratory for Information Science and Technology\\
        Department of Electronic Engineering, Tsinghua University, Beijing 100084, China \\
Email: zhaotc13@mails.tsinghua.edu.cn, sheng.zhou@tsinghua.edu.cn, guo-xy11@mails.tsinghua.edu.cn\\
zhaoyun12@mails.tsinghua.edu.cn, niuzhs@tsinghua.edu.cn\\}
}

\maketitle

\begin{abstract}
With the proliferation of mobile applications, Mobile Cloud Computing (MCC) has been proposed to help mobile devices save energy and improve computation performance.
To further improve the quality of service (QoS) of MCC, cloud servers can be deployed locally so that the latency is decreased.
However, the computational resource of the local cloud is generally limited.
In this paper, we design a threshold-based policy to improve the QoS of MCC by cooperation of the local cloud and Internet cloud resources, which takes the advantages of low latency of the local cloud and abundant computational resources of the Internet cloud simultaneously.
This policy also applies a priority queue in terms of delay requirements of applications.
The optimal thresholds depending on the traffic load is obtained via a proposed algorithm.
Numerical results show that the QoS can be greatly enhanced with the assistance of Internet cloud when the local cloud is overloaded.
Better QoS is achieved if the local cloud orders tasks according to their delay requirements, where delay-sensitive applications are executed ahead of delay-tolerant applications.
Moreover, the optimal thresholds of the policy have a sound impact on the QoS of the system.

\end{abstract}


%
\IEEEpeerreviewmaketitle

\section{Introduction}
The amount of mobile applications increased dramatically in recent years.
By April 2015, Android users have accessed to more than $1.5$ million applications \cite{APP}.
This trend enhances the Quality of Service (QoS) of mobile devices, but the energy consumption is also increased.
In fact, the plethora of applications caused heavy energy consumption, which significantly reduces the battery life of smart phones.
Remote execution is a possible way to help smart phones save energy.
By offloading energy-intensive tasks to resource-rich servers, battery life of mobile devices can be significantly improved \cite{MCCsaveenergy}.
Based on Mobile Cloud Computing (MCC), some platforms are designed, e.g., MAUI \cite{MAUI}, CloneCloud \cite{CloneCloud}.

Although remote execution is very prominent in terms of energy saving, it brings challenges to guarantee latency.
Delay is a very important QoS requirement from mobile users \cite{latency}.
However, if an application is offloaded to a remote centralized cloud server, the delay requirement can hardly be satisfied because of the long transmission delay over the Internet \cite{MCCdelay1} \cite{MCCdelay2}.
Cloudlet \cite{cloudlet} is proposed to deploy some local cloud servers, so that delay requirement can be met.
Some specific architectures focusing on technological details are designed then, such as FemtoCloud \cite{Femto}, CONCERT \cite{CONCERT}.
In the proposed architecture, each local cloud serves mobile users of several nearby cells, which indicates that local cloud should be deployed densely with a large number.
Our earlier work \cite{multiplexing} studied and analysed how much computational resources need to be deployed in a cloud so that they can be used efficiently.
It is concluded that if computational resources exceeds a threshold, the extra resources only provide marginal gain.
Based on the analysis, computational resources of each local cloud should be deployed reasonably so as to balance the cost and QoS.\par

Thus, although local cloud is beneficial in terms of transmission delay, its computational resources is relatively limited.
An architecture to associate cloudlets is proposed in \cite{zs2}, which takes the advantage of cloudlets cooperation to overcome computational resource limitation of a single cloudlet.
Yet computational resource of cloudlets still has a limitation, so that system performance might degrade when the traffic load is high.
Remote cloud has sufficient computational resources, and it can cooperate with the local cloud to achieve better QoS.
Load sharing between the local cloud and remote cloud is studied in \cite{zs3}, which is optimized in terms of average response time and energy consumption.
But each application has a delay requirement bound, and it is not practical to evaluate the performance of this kind of traffic by average delay.\par

\begin{figure}[h]
  \centering
  \includegraphics[width=8cm]{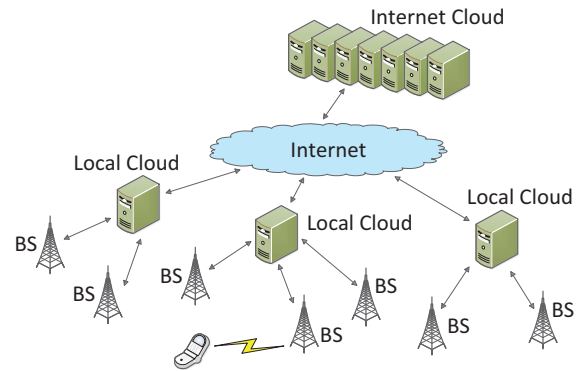}
  \caption{A cellular network with local cloud and Internet cloud.}
\end{figure}


As application data needs to be transmitted to remote cloud by Internet, we define remote cloud as Internet cloud.
The transmission delay of remote execution can be quite long and the delay jitter is generally large.
One simple intuition is that, delay-sensitive applications should be executed in the local cloud, while delay-tolerant tasks can be offloaded to Internet cloud when traffic load is heavy.
By the cooperation of local cloud and Internet cloud, computational resources can be used efficiently and QoS requirements can be satisfied.\par

We illustrate our idea in Fig. 1.
Resource-constraint local cloud is near to mobile users, while resource-abundant Internet cloud is remotely located.
Mobile users access the local cloud through wireless communication and fronthaul transmission, while Internet transmission is in addition to them if users access the Internet cloud.
As local cloud has limited computational resources, some arriving tasks might need to wait longer to be served.
To enhance the QoS, we design a scheduling policy so that delay requirements of more users are satisfied.
The policy cooperatively schedules the resources in the local cloud and Internet cloud.
When the traffic load of the local cloud is above a certain threshold, delay-tolerant applications have to be offloaded to the Internet cloud in order to leave more local computational resources for delay-sensitive applications.
To further enhance the QoS, We model the local cloud as a priority queue system.
For delay-sensitive applications, they are labeled with higher priority and will be executed ahead of delay-tolerant applications.

The rest of the paper is organized as follows.
Section \Rmnum{2} introduces the system model and gives the problem formulation.
Section \Rmnum{3} proposes the scheduling policy and analyses its performance.
Section \Rmnum{4} shows numerical results to evaluate the proposed policy.
The paper is concluded in Section \Rmnum{5}.\\

\section{System model and problem formulation}

The system model is shown in Fig. 2.
Mobile users offload their applications to cloud servers.
The data is firstly transmitted to a scheduler which is located in the local cloud.
The scheduler decides whether to execute the application in the local cloud or send it to the Internet cloud.
The application is then executed in one of the clouds.
As soon as the execution is completed, the result will be sent back to the scheduler, and it is finally fed back to mobile users.
In this model, wireless transmission and fronthaul transmission between users and the local cloud is needed no matter where to execute the application, whose delay is considered as a small constant $\tau$.

\begin{figure}[h]
  \centering
  \includegraphics[width=8.5cm]{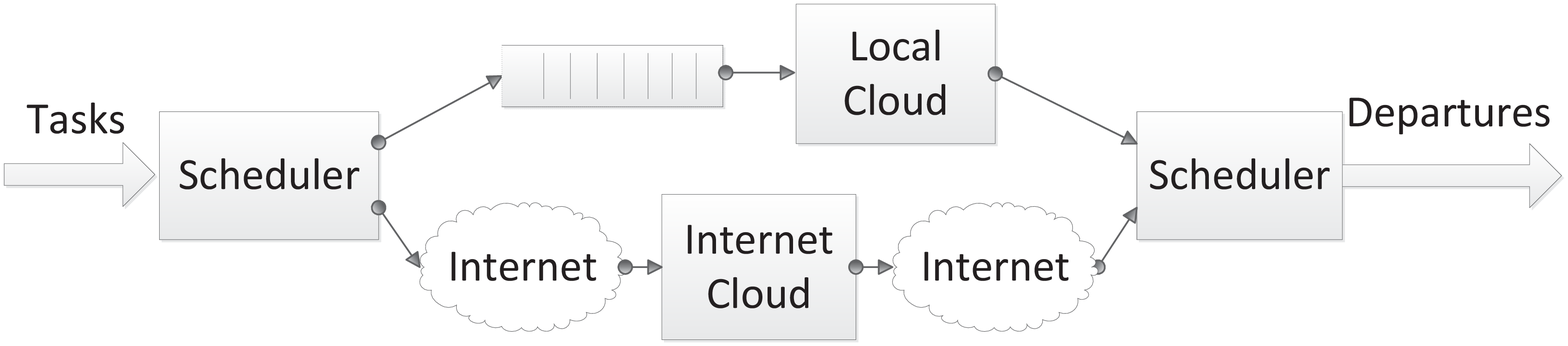}\\
  \caption{System model.}
\end{figure}

Each application offloaded from mobile devices is considered as an arriving task, and each task has a delay constraint.
The task is successfully completed if it is executed within the constraint.
We design a resource allocation policy to optimize the success probability.\\

\subsection{Tasks}
Assume that there are $N$ types of tasks with different delay requirements.
Tasks of type $i$ arrive at the scheduler in a Poisson process with rate $\lambda_i$, and they are executed by either the local cloud or the Internet cloud.
Each of them has a system delay constraint $T_i (i=1,...,N)$ which is the delay requirement minus $\tau$.
These tasks request exponentially distributed service time with parameter $\mu$.
Rank the priority of these tasks according to delay requirements, and a delay requirement vector $\boldsymbol{T}=(T_1,T_2,...,T_N)$ is given:

\begin{equation}\label{1}
  T_1 \leq T_2 \leq ... \leq T_N
\end{equation}\par

\subsection{Local Cloud}

Assume that there are $C$ virtual machines working in the local cloud.
Each virtual machine can be seen as a single server.
In the queuing system, we assign different priorities to tasks of different delay requirements, where a task of smaller delay requirement has a higher priority.
The system forms a nonpreemptive priority queue, where tasks being executed will not be interrupted when a higher priority task comes.
Meanwhile, the system has a finite buffer for each type of tasks.
Accordingly, the local cloud is modeled as an $M/M/C$ system with modified preemptive priorities.
In our model, the total delay is composed of two parts, which includes the queuing delay and execution delay.
The local cloud model is shown in Fig. 3.

\begin{figure}[h]
  \centering
  \includegraphics[width=8.5cm]{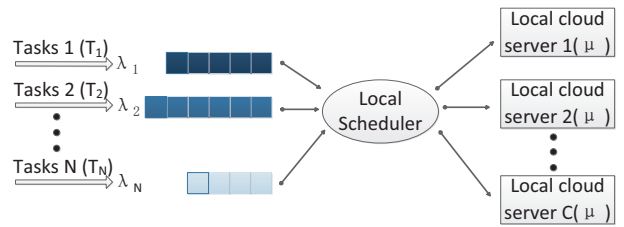}\\
  \caption{Local Cloud model.}
\end{figure}

\subsection{Internet Cloud}

The mean delay of Internet transmission is assumed to be long and so does the delay jitter.
Thus, Internet cloud is by no means a good choice in terms of delay requirement.
However, if the number of coming tasks exceeds the service capacity of the local cloud, Internet cloud should be used to help improve the probability that the delay requirement is met.
As the Internet cloud is abundant in computational resources, we assume that the execution delay can be ignored compared with the transmission delay.\par
Some related works model Internet transmission delay, and we adopt the model of reference \cite{Internetdelay}.
They propose that $\varphi_1(t)$ is the router delay distribution and $\varphi_2(t)$ is the queuing delay distribution, the Internet transmission delay distribution is

\begin{equation}\label{1}
  \varphi(t)=p\varphi_1(t)+q\varphi_1(t) * \varphi_2(t)
\end{equation}\par

\subsection{Optimization Objective}

The optimization objective is the probability that an arriving task is successfully executed within its delay constraint $T_i$.
The constraint of the system is the limited computational resource of the local cloud.
The maximum number of servers can be used in the local cloud is $C$.
The objective is $P_\text{success}(t<T_i|C)$, where $t$ indicates the time consumption.
We design a scheduling policy to optimize the objective by deciding when and where to execute the arrival tasks.

\section{Policy Designment and Performance Optimization}

Assume that the arrival rates and service rate of all types of tasks are given, which are $\boldsymbol{\lambda}=(\lambda_1,...,\lambda_N)$ and $\mu$ separately.
Define the vector $\boldsymbol{S}=(s,l_1,l_2,...,l_N)$ as the state of the queuing system.
The first parameter $s$ denotes the number of busy servers in the local cloud, and $l_i(i=1,2,...,N)$ denotes the number of tasks of priority $i$ waiting in the queue.\par
We design a threshold based policy to cooperate the local cloud and Internet cloud, which schedules tasks according to their delay requirements.
The policy is shown as follows.

\textbf{Priority-based Cooperation policy}:
If there is at least one empty server, a task is executed in the local cloud as soon as it arrives.
Otherwise, the arrived task of priority $i$ waits in the queue of its own type.
If a task departures the system, the empty server will execute a waiting task of the highest priority if any.
The priority-based sub-queues are illustrated in Fig. 3 with different colors.
For tasks of priority $i$ or higher, a buffer threshold $B_i$ is set to contain them.
If the buffer $B_i$ is full, the coming task of priority $i$ or higher is offloaded to the Internet cloud.
The buffer thresholds vector $\boldsymbol{B}=(B_1,...,B_N)$ is derived accordingly.\par

The intuition of the policy designment is explained as follows.
Firstly, tasks of higher priorities have shorter delay requirements.
Tasks of lower priorities could wait in the queue until higher-priority tasks have been executed.
Secondly, the sojourn time distribution is determined by queue length.
By optimizing threshold $B_i$, the success probability of tasks can be enhanced.
Finally, if a priority-$i$ task is the last one in the queue and the queue length equals to the threshold, the policy do not permit a higher-priority task to enter the queue.
Otherwise, the priority-$i$ task may suffer from low success probability because of a burst of higher-priority traffic.\par

To further evaluate the performance of our policy, we design some classical policies for comparison, which are shown as follows.

\textbf{Local Cloud policy}:
The tasks are executed only in the local cloud.
The system is $M/M/C$ with preemptive priority.\par
\textbf{Greedy policy}:
The coming task chooses the better one between the local cloud and Internet cloud so that it will have a higher success probability.\par
\textbf{FCFS-based Cooperation policy}:
Local cloud is a $M/M/C$ system with First Come First Serve (FCFS) queue.
The coming task is offloaded to either the local cloud or the Internet cloud by comparing current queue length with a threshold.\par
\textbf{Non-buffer policy}:
If all local cloud servers are being used, the coming task will be offloaded to the Internet cloud.

\subsection{Stationary Distribution}

The queuing system is a $N$-dimension Markov chain.
We can get stationary distribution by formulating and solving global balance equation.
Define $L_i=\sum_{j=1}^{i} l_j$ and $\Lambda_i=\sum_{j=1}^{i} \lambda_j$, $\rho_i=\frac{\Lambda_i}{\mu}$.
For $L_N=0$,
\begin{equation}\label{1}
\begin{aligned}
  & {(\Lambda_N+s\mu)p(s,0,0,...,0)=}  \\ &{(s+1)\mu p(s+1,0,0,...,0)+\Lambda_N p(s-1,0,0,...,0)}
  \end{aligned}
\end{equation}\par

For $0 < L_i < B_i$,
\begin{equation}\label{1}
\begin{aligned}
 & (\Lambda_N + C\mu)p(C,l_1,l_2,...,l_N) = \\ & \sum_{j=1}^{N}{ \lambda_j p(C,l_1,...,l_j-1,...,l_N)} \\ & +\sum_{j=1}^{M}C\mu p(C,l_1,...,l_j+1,...,l_N)
   \end{aligned}
\end{equation}\par
Here, $M$ is the type of tasks of the highest priority in the queue, and the queuing system state is $(C,0,...,0,l_M,l_{M+1},...,l_N)$.

For the maximum $i$ satisfying $L_i = B_i$,
\begin{equation}\label{1}
\begin{aligned}
 &(\Lambda_N-\Lambda_i + C\mu)p(C,l_1,l_2,...,l_N) = \\ & \sum_{j=1}^{N}{ \lambda_j p(C,l_1,...,l_j-1,...,l_N)}
   \end{aligned}
\end{equation}\par

This is a $N$-dimension Markov chain, and it has only one stationary distribution.\\
\emph{Proof}: States $(s,0,...,0)$ and states $(C,0,...,0)$ communicate with each other, which is denoted as $(s,0,...,0) \leftrightarrow{} (C,0,...,0)$.
The states also have the following relations.
\begin{equation}\label{1}
  (C,l_1,...,l_i,...,l_N) \leftrightarrow{} (C,0,l_2,...,l_i,...,l_N)
\end{equation}\\
\begin{equation}\label{1}
  (C,0,...,0,l_i,...,l_N) \leftrightarrow{} (C,0,,...,0,l_{i+1},...,l_N)
\end{equation}

Thus, all states communicate with each other, which indicates that the Markov chain is irreducible.
The Markov chain has a stationary distribution and no other stationary distribution exists \cite{irreducible}.\par

If $N$=2, the 2-dimension Markov chain is shown in Fig. 4.
The states $(i)$ which are below the dashed line represent the number of busy servers, where queue is empty.
The states $(l_i,l_j)$ which are above the dashed line represent queue lengths of different types of tasks, where all servers are busy.
The states $(l_i,l_j)$ whose $l_i=B_1$ or $l_i+l_j=B_2$ indicate that the buffer is full, and arriving tasks will be offloaded to the Internet cloud.

\begin{figure}[h]
  \centering
  \includegraphics[width=8.5cm]{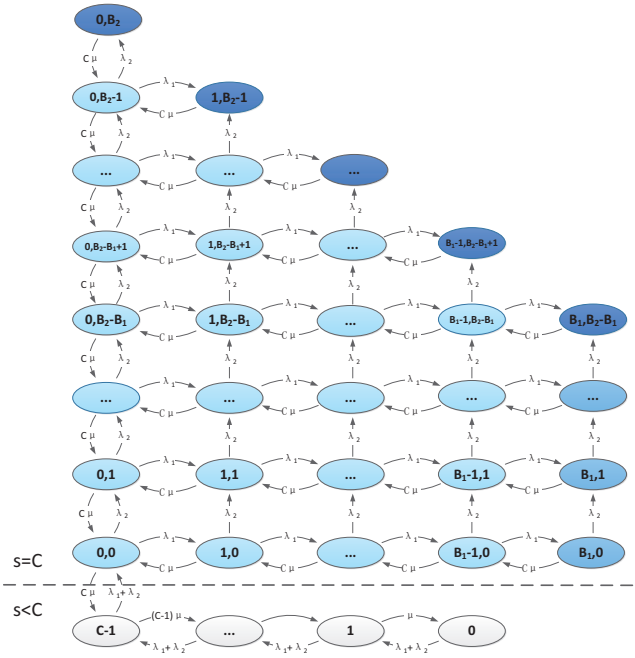}\\
  \caption{Markov chain for 2-priorities system.}
\end{figure}

\subsection{Sojourn Time Distribution}

System state is $\boldsymbol{S}=(s,l_1,...,l_N)$.
If $s \leq C$, a task is served as soon as it arrives at the queuing system.
The sojourn time equals to the service time, which is exponentially distributed.
The pdf of sojourn time is:

\begin{equation}\label{1}
  p_\text{st}(t|\boldsymbol{S})=\mu e^{\mu t} \quad(t \geq 0)
\end{equation}\par

If $s > C$, all servers are being used when a task arrives at the queuing system.
The task has to wait in the queue or served by the Internet cloud.
If the task is served by the local cloud, it needs to wait in the queue until it can be served.
The sojourn time consists of waiting time and service time.\par

Assume that the priority of the arriving task is $i$.
The distribution of waiting time for the task $w(t)$ is $L_i+1$ fold convolution of $f(t)$ which is the busy period of a $C$-server system serving higher-than-$i$ priority tasks \cite{priorityqueuewaitingtime}. The probability density function $f(t)$ and its Laplace-Stieltjes transform are as follow \cite{busyperiod}.

\begin{equation}\label{1}
  f(t)=\frac{1}{t\sqrt{\rho_{i-1}}}e^{-(\Lambda_{i-1}+\mu)t}I_1(2t\sqrt{\Lambda_{i-1}\mu})\quad(t \geq 0)
\end{equation}\par
\begin{equation}\label{1}
\begin{aligned}
  &\bar{F}(s)= \\& (s+\Lambda_{i-1}+C\mu-\sqrt{(s+\Lambda_{i-1}+C\mu)^2-4c\mu\Lambda_{i-1}})/2\Lambda_{i-1}
  \end{aligned}
\end{equation}\par

$I_1 $ is a modified Bessel function of the first kind. Thus, the distribution of waiting time $w(t)$ and its Laplace-Stieltjes transform is derived.

\begin{equation}\label{1}
  w(t|\boldsymbol{S})=f(t)*f(t)*...*f(t)
\end{equation}\par
\begin{equation}\label{1}
  \bar{W}(s|\boldsymbol{S})=\bar{F}(s)^{L_i+1} 
\end{equation}\par

The sojourn time distribution $p_\text{st}(t|\boldsymbol{S})$ is the convolution of waiting time and service time.
\begin{equation}\label{1}
  p_\text{st}(t|\boldsymbol{S})=w(t|\boldsymbol{S})*\mu e^{\mu t}
\end{equation}\par

Given the state $\boldsymbol{S}$ of the system, success probability is
\begin{equation}\label{1}
  P_\text{success}(t \leq T_i|\boldsymbol{S})=\int_{0}^{T_i} p_\text{st}(t|\boldsymbol{S}) dt \qquad
\end{equation}\par

If the task is served by the Internet cloud, the sojourn time distribution is modeled as an empirical distribution $P_\text{I}(t)$ which is given in (2).\\

\subsection{Success Probability}

Assume that $\boldsymbol{\lambda},\mu $ and $ \boldsymbol{B} $ are given, the success probability of priority-$i$ tasks is calculated as:

\begin{equation}\label{1}
\begin{aligned}
 &P_\text{success}(i|\boldsymbol{\lambda},\mu,\boldsymbol{B})= P(t \leq T_i|\boldsymbol{\lambda},\mu,\boldsymbol{B})= \\ & \sum_{L_j<B_j}{P_\text{st}(t \leq T_i|\boldsymbol{S})p(s,l_1,l_2,...,l_N)} \\ & +\sum_{L_j=B_j}{P_\text{I}(t \leq T_i)p(s,l_1,l_2,...,l_N)}
\end{aligned}
\end{equation}\par

Note that the $P_\text{st}(t \leq T_i|\boldsymbol{S})$ in equation (15) is related to $\boldsymbol{\lambda},\mu $ and $ \boldsymbol{B}$, which is given in equation (14).

The total success probability is
\begin{equation}\label{eq:success probability}
\begin{aligned}
 P_\text{success}(\boldsymbol{\lambda},\mu,\boldsymbol{B})=\frac{\sum_{i=1}^{N}{\lambda_i P_\text{success}(i|\boldsymbol{\lambda},\mu,\boldsymbol{B})}}{\Lambda_N}
\end{aligned}
\end{equation}\par

\subsection{Local Optimal Thresholds}

Search algorithm can be used to get the optimal thresholds vector $(B_1,...,B_N)$, so that the success probability is maximized.
But the complexity of search algorithm might be quite high.
Here, we give a low-complexity recursive algorithm to get the local optimal thresholds, which is shown in the Algorithm 1.
Firstly, make $(B_1,...,B_N)=(0,...,0)$.
Secondly, continually increase $B_N$ by 1 until $P_\text{success}$ begin to decrease, and a local optimal $B_N$ is derived given that $(B_1,...,B_{N-1})=(0,...,0)$.
Next, make $(B_0,...,B_{i-1})=(0,...,0)$.
Increase $B_i$ by 1 each time to get optimal $(B_{i+1},...,B_N)$, and stop adding $B_i$ until $P_\text{success}$ decreases.
Repeat the previous step to get the buffer thresholds $(B_1,...,B_N)$.
This algorithm gives local optimal thresholds, while search algorithm is optimal globally.
In the simulation scenarios, numerical results show that the thresholds of our algorithm equal to the thresholds derived by search algorithm.

\renewcommand{\algorithmicrequire}{\textbf{Input:}}
\renewcommand{\algorithmicensure}{\textbf{Output:}}

\begin{algorithm}
\caption{Find local optimal thresholds}
\begin{algorithmic}[1]

\Require $\boldsymbol{\lambda}=(\lambda_1,...,\lambda_N),\mu$
\Ensure $\boldsymbol{B}=(B_1,...,B_N)$

\State $\boldsymbol{B} \gets (0,...,0)$
\State $\boldsymbol{B} \gets $ \Call{FindOptimalThreshold} {$1, N, \boldsymbol{B}$}

\Statex
\Procedure {FindOptimalThreshold}{$i$, $N$, $\boldsymbol{B}$}
\For{$k\gets i \text{ to } N$}
\State $B_k \gets B_{i-1}$
\EndFor
\If {$i = N$}
\State $P_\text{Success1} \gets P_\text{success}(\boldsymbol{\lambda},\mu,\boldsymbol{B})$
\State $P_\text{Success2} \gets P_\text{Success1}$
\While{$ P_\text{Success1} \leq P_\text{Success2}$ }
\State $B_N \gets B_N+1$
\State $P_\text{Success1} \gets P_\text{Success2}$
\State $P_\text{Success2} \gets P_\text{success}(\boldsymbol{\lambda},\mu,\boldsymbol{B})$
\EndWhile
\State $B_N \gets B_N-1$
\Else
\State $\boldsymbol{B} \gets $ \Call{FindOptimalThreshold} {$i+1, N, \boldsymbol{B}$}
\State $P_\text{Success1} \gets P_\text{success}(\boldsymbol{\lambda},\mu,\boldsymbol{B})$
\State $P_\text{Success2} \gets P_\text{Success1}$
\While{$ P_\text{Success1} \leq P_\text{Success2}$ }
\State $B_i \gets B_i+1$
\State $\boldsymbol{B} \gets $ \Call{FindOptimalThreshold} {$i+1, N, \boldsymbol{B}$}
\State $P_\text{Success1} \gets P_\text{Success2}$
\State $P_\text{Success2} \gets P_\text{success}(\boldsymbol{\lambda},\mu,\boldsymbol{B})$

\EndWhile
\State $B_i \gets B_i-1$
\State $\boldsymbol{B} \gets $ \Call{FindOptimalThreshold} {$i+1, N, \boldsymbol{B}$}
\EndIf
\State \textbf{return} $\boldsymbol{B}$
\EndProcedure

\end{algorithmic}
\end{algorithm}

\section{Numerical results}

We evaluate the proposed priority-based cooperation policy by comparing it with other policies stated previously.
In the evaluation, two types of tasks are considered, which are delay-sensitive tasks and delay-tolerant tasks separately.
We assume the parameters of the system as follows.
Delay requirement of delay-sensitive tasks is 50 milliseconds, and delay requirement of delay-tolerant tasks is 300 milliseconds.
The Internet delay is modeled in (2), whose mean delay is 200 milliseconds.
For the local cloud server, its mean service time is 10 milliseconds.\par

Fig. 5 shows the comparison between priority-based cooperation policy and local cloud policy.
When traffic load is low, the local cloud has enough computational resources to execute arriving tasks and most users can complete their tasks within delay requirements.
However, the success probability decreases dramatically with the increasing of arrival rate.
In fact, most users have to wait in the queue when traffic load is heavy, which leads to poor QoS.
In this case, cooperation of the local cloud and Internet cloud is quite necessary.
By offloading delay-tolerant tasks to Internet cloud, much more mobile users can have their applications completed successfully.
In our model, more than 20\% success probability of tasks can be enhanced by cooperation of the local cloud and Internet cloud when traffic load is heavy.

\begin{figure}[h]
  \centering
  \includegraphics[width=8.5cm]{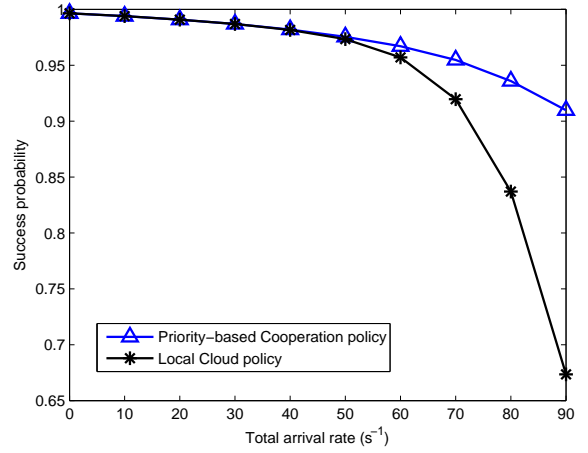}\\
  \caption{Success probability vs. Arrival rate for Priority-based Cooperation policy and Local Cloud policy.}
\end{figure}

Fig. 6 shows the comparison between priority policy and non-priority policies.
A single user can achieve higher QoS by greedy policy which maximizes his success probability.
However, optimization of a user becomes a burden for the system, because it results in longer mean waiting time.
In fact, local optimum is by no means global optimum.
The scheduling policy needs to be designed globally so that higher success probability for total users can be achieved.
For the FCFS-based cooperation policy, it fully utilizes the computational resources of the local cloud and the Internet cloud and its performance is quite good.
But higher QoS can be realized by considering priorities of tasks.
Results shows that the priority policy is better than the FCFS policy to a certain extent.
In our model, it results in a 5\% success probability improvement if the policy considers priorities of tasks.
Non-buffer policy only makes decisions according to the state of servers.
It fails to utilize the buffer to make future plans and further improve the QoS, which results in a bad QoS performance.

\begin{figure}[h]
  \centering
  \includegraphics[width=8.5cm]{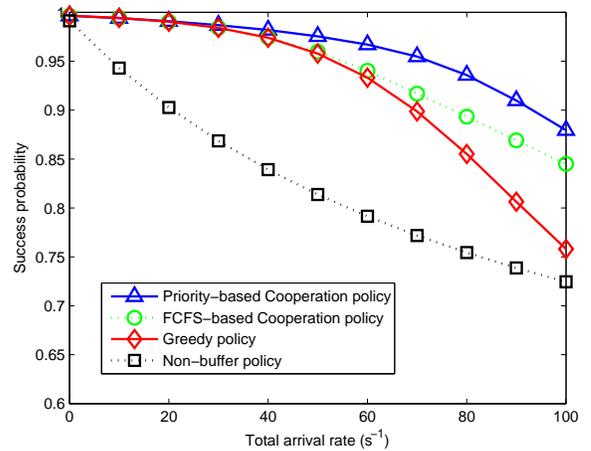}\\
  \caption{Success probability vs. Arrival rate for Priority-based Cooperation policy and Non-Priority Cooperation policies.}
\end{figure}

Fig. 7 gives the optimal thresholds of our proposed policy.
When traffic load is low, threshold $B_2$ is a small value.
In this condition, extra buffer is not needed and a small threshold will achieve the optimal success probability.
With the increase of arrival rate, a larger buffer threshold is essential to hold more tasks in the local cloud.
However, when traffic load is heavy, the buffer threshold should decrease, because the queue will always be full and long queue length leads to large waiting time.
Threshold $B_1$ decreases with the increase of arrival rate, which leads to waiting time reduction of all types of tasks.

\begin{figure}[h]
  \centering
  \includegraphics[width=8.5cm]{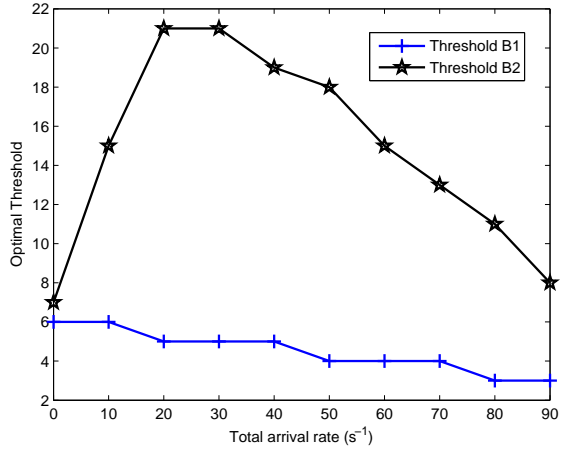}\\
  \caption{Optimal Thresholds vs. Arrival rate.}
\end{figure}

\section{Conclusion}

In this article, we have improved the QoS of MCC users by designing a scheduling scheme to realize the cooperation between the local cloud and the Internet cloud.
We firstly classify applications according to their delay requirements, and give higher priority to applications with shorter delay requirements.
Then, we design a threshold-based policy to cooperatively scheduling the local cloud and the Internet cloud, so that the QoS is dramatically improved.
By optimizing the thresholds, probability that tasks can be executed within their delay requirements is maximized.
We further give an recursive algorithm to get the optimal thresholds with low computation complexity.
Numerical results reveal that:
1) Limited computational resources of the local cloud greatly influences the QoS when the traffic load is high, and Internet cloud is needed to improve QoS.
By cooperation of the local cloud and Internet cloud, probability that a task is completed within its delay requirement can be improved by 20\%.
2) The QoS can be further improved by 5\% via a priority scheme which executes delay-sensitive tasks ahead of delay-tolerant tasks.
3) Optimal buffer thresholds are tightly related to the traffic load. As the traffic load increases, a larger buffer threshold is needed to hold more tasks. But thresholds should decrease to guarantee a small waiting time when traffic load is heavy.\par


\section*{Acknowledgment}

This work is sponsored in part by the National Basic Research Program of China (973 Program: No. 2012CB316001), the National Science Foundation of China (NSFC) under grant No. 61201191, No. 61322111, No. 61321061, No. 61401250, and No. 61461136004, and Hitachi Ltd.

\emph{}


%

\end{document}